\documentstyle[aps,amsfonts,preprint]{revtex}

\newtheorem{observation}{Observation}

\begin{document}
\title{
  {THE CONSTRUCTION OF SPINOR FIELDS}
                {ON}
  {MANIFOLDS WITH SMOOTH DEGENERATE METRICS}
  }

\author{
J\"org Schray{\footnotemark },
Tevian Dray{\footnotemark },
Corinne A. Manogue{\footnotemark },
Robin W. Tucker{\footnotemark },
Charles Wang{\footnotemark }
}
\address{School of Physics and Chemistry, Lancaster University,
		Lancaster LA1 4YB, UK
}

\maketitle

\begin{abstract}

We examine some of the subtleties inherent in formulating a theory of
spinors on a manifold with a smooth degenerate metric. We concentrate on the
case
where the metric is singular on a hypersurface that partitions the manifold
into Lorentzian and Euclidean domains.  We introduce the notion of a
complex spinor fibration  to make precise the meaning of continuity of a
spinor field and give an expression for the components of a local
spinor connection that is valid in the absence of a
frame of local orthonormal vectors.
These considerations enable one to construct a Dirac equation for the
discussion of the behavior of spinors in the vicinity of the metric
degeneracy. We conclude that the theory contains more freedom than the
spacetime Dirac theory and we discuss some of the implications of this for
the continuity of conserved currents.

\end{abstract}

\ifpreprintsty
\bigskip\smallskip
{\def\labelitemi{}
PACS:
\begin{itemize}
\item
\shortstack[l]
{\noalign{\vskip 6pt}
04.20.Gz Spacetime topology, causal structure, spinor structure\cr
\noalign{\vskip 6pt}
11.90.+t Other topics in general field and particle theory\cr
\noalign{\vskip 6pt}
03.65.Pm Relativistic wave equations\cr
\noalign{\vskip 6pt}
03.50.Kk Other special classical field theories\cr
}
\end{itemize}
}
\fi

\section{Introduction}

The interest in the influence of topology on physics is an old one. In
recent times there has also been considerable debate on the influence of the
geometrical structure of spacetime that may accompany a change in its
overall  topology.
This has been partly motivated by the implications of the semi-classical
theory of quantum gravity and partly by the interest in field theories on
background spacetimes with interesting topologies.  Further motivation
arises from string theories in which string interactions  arise from the
topology of world sheets. In all these approaches fundamental assumptions
about the signature of the spacetime metric are required. Such assumptions
dictate the detailed behavior of both the causal structure of the theory
and the selection rules for topology change.
In the context of classical theory
there are powerful constraints on the nature of such changes on manifolds
with a global Lorentzian signature and a spinor structure
\cite{Gibbons Hawking}.
To escape such
constraints a number of authors have contemplated  geometries in which the
metric is allowed to become degenerate, particularly on hypersurfaces that
partition the manifold into Lorentzian and Euclidean regions. Despite the
obvious implications for  causality there have been serious attempts to
follow the consequences for physics associated with signature changing
metrics.
Despite the absence of a rigorous theory of second quantized fields on such
a background, in \cite{Dray et al.} it was
suggested that a quantized scalar field
 could exhibit spontaneous particle production even in the
absence of gravitational curvature. This result relied on certain natural
linear
boundary conditions that were imposed on the scalar field at the
hypersurface of signature change. Since there is no continuous
orthonormal coframe in the presence of metric degeneracy
and the field equations are themselves dependent on the metric one must rely
on a prescribed differential structure in order to define the necessary
limits of the gradients of the scalar field in the vicinity of the metric
degeneracy. In practice this means one can always rely on a local
coordinate coframe to effect one's calculations. Furthermore
the differentiability class of all tensor fields  is defined with respect
to the differentiability of their components in an arbitrary coordinate
(co-)frame independent of any metric structure.

Since matter in flat Lorentzian spacetime is also  described in terms of
various representations of the Lorentzian SPIN group it is natural to try
and extend these considerations to the behavior of spinor fields on
manifolds with a degenerate metric.  In particular one may wish to
formulate a dynamical theory of spinor fields and deduce from their field
equations a class of natural boundary conditions at the hypersurface of
signature change. However a number of interesting problems then
arise that have no counterpart in the theory of tensor fields.
The most obvious is that the dimensionality of the real irreducible SPIN
representations is signature dependent so that it becomes meaningless to
try and match spinor fields belonging to representations with different
dimensions. If one persists with the search for matching conditions one must
in general consider complex representations.

In a smooth local basis of spinor fields one can define the differentiability
class of the components of a spinor field. Such a basis is a basis for a
module carrying representations of the SPIN group, which is a double cover of
the $SO(p,q)$ group associated with the signature of the underlying metric on
the manifold.
Clearly  this procedure will fail at the hypersurface where the
signature changes, since the SPIN groups differ across the hypersurface.
In order to define continuous spinor fields on a neighborhood crossing the
hypersurface, alternatives to the traditional reliance on lifting orthonormal
frames to spinor frames must be pursued.
Of necessity one must expect some
arbitrariness in defining the notion of a continuous spinor field in the
presence of signature change.

It is natural to subject local spinor fields to the appropriate
Dirac equation in
regions where the metric is non-degenerate.
In such regions the  conventional  Dirac operator can be defined in terms
of a spinor covariant derivative that is designed to satisfy the natural
Leibniz rules on products of tenors and spinors. In this manner it can be
made compatible with the natural linear connection on tensors. A unique
Levi-Civita
tensor covariant derivative is determined completely by the metric tensor.
When this metric is non-degenerate one can exploit the existence of local
orthonormal frames to uniquely
fix the spinor connection that determines the spinor
covariant derivative. It is important to stress that it is only the
{\it existence} of a class of orthonormal frames that is necessary to effect
this determination, since it provides a reference frame for
normalization.
The SPIN connection so defined is then compatible with a SPIN invariant
inner product on spinors.
If one attempts to define a spinor connection in the absence of a class of
orthonormal frames then one must recognize the inherent arbitrariness
that cannot be removed by normalization. Since we are interested in
subjecting our spinor fields to the appropriate Dirac equation in regions
where the metric is regular we  must
accommodate this freedom in the spinor connection if we wish to discuss the
matching of spinor solutions at the hypersurface of degeneracy.

Little attention has been devoted to the formulation of spinor fields on
spaces with degenerate metrics.
Romano \cite{Romano} recognized that the choice of spinor equation was not
straightforward.  His analysis was restricted to the case of a discontinuous
change of signature, whereas in this article we restrict ourselves instead to
the case of continuous degenerate metrics.
It is our purpose to examine
the essential arbitrariness inherent in a formulation of spinor theory on
manifolds with such metrics.

In section II we offer a definition of complex  spinors in terms
of a {\it spinor fibration}
over a manifold. Although our construction relies on
the representation theory of Clifford algebras, we have translated our
arguments into the traditional language of $\gamma$ matrices. The essential
novelty is that these are matrix representations of  a set of
coordinate vector fields
that constitute a frame in the vicinity of the metric degeneracy. The
representation structure is explicitly presented in terms of degenerate
metrics in two and four dimensions.

Having defined the notion of spinor continuity in terms of a spinor
fibration, we turn to the notion of the spinor covariant derivative
in section III. We show how this can be determined to be both compatible with
a SPIN invariant inner product and to commute with the complex structure
(``charge conjugation''). In section IV we write down and solve the two
dimensional Dirac equation written in terms of this spin connection,
making explicit the dependence of the singularity structure of these
solutions on both the spin metric and the metric on the underlying manifold.
We conclude with a brief discussion of the $U(1)$ currents associated
with these solutions and offer some speculations on alternative approaches.

\section{Spinors}

In $n = 2 \, m$ dimensions we consider the manifold $M={\Bbb R}^{2m}$ with
metric
\begin{equation}
{\rm g} = {h(t)}\, dt \otimes dt + \hat{\rm g}_{ij}(\vec x)\,dx^i \otimes dx^j
\label{eq:metric4}
\end{equation}
in a chart $(t,x^i) = (x^\mu)$, $i = 1$, \ldots, $n-1$, where $\hat{\rm g}$
is assumed to be positive definite.
${h}$ is a smooth function which may have zeroes (at most countably many
that are nowhere dense).  However, we require
that ${h}$ changes sign at zeroes of ${h}$.
None of the crucial steps of the development below rely on the topological
triviality of this particular manifold.
Although the discussion applies to complex spinors on any even dimensional
manifold with signature change, we will pay particular attention to the cases
$n=4$ and $n=2$.

Kossowski and Kriele have shown \cite{KK} under fairly general conditions
that, at any zero of ${h}$ where $\dot{h} \not= 0$, one can switch to
coordinates $(t',x^i)$ in a neighborhood of the zero such that ${h(t)} \, dt^2
= t' \, d{t'}^2$.  However, the precise nature of the signature change is not
of importance within the scope of this article.

To define Dirac spinors on a manifold $M$ of constant signature one usually
\cite{Benn and Tucker}
considers local irreducible representations $\bbox\gamma$ of the complex
Clifford algebra bundle
\begin{equation}
{\cal C \it l}(M)
= \mathop{\bigcup}\limits_{p \in M} {\cal C \it l}(T_pM,{\rm g}_p),
\end{equation}
i.e, $\bbox\gamma$ is a fiber preserving homomorphism
\begin{equation}
\bbox\gamma: \pi^{-1}(U) \subset {\cal C \it l}(M) \to M_k({\Bbb C}) \times U,
\end{equation}
where $\pi:{\cal C \it l}(M) \to M$ is the bundle projection, $U$ is an open
subset of $M$, and $M_k({Bbb C})$ is the set of complex $k\times k$ matrices
($k = 2^m$).
We will assume for now that $\bbox\gamma$ is at least continuous.
If the representation $\bbox\gamma$ is also faithful, which is the case for
even dimension of $M$, then $\bbox\gamma$ is just a local trivialization of
${\cal C \it l}(M)$.
In particular, for vector fields $X$ and $Y$,  $\bbox\gamma$ satisfies
\begin{equation}
\{\bbox\gamma(X),\bbox\gamma(Y)\} = 2\, {\rm g}(X,Y)\bbox 1.
\label{eq:rep}
\end{equation}
With respect to a local coordinate chart $\bbox\gamma$ is
given by its components
\begin{equation}
\gamma_\mu := \bbox\gamma(\partial_\mu).
\end{equation}
(Note that we use bold-faced $\bbox\gamma$ for the representation map, and
light-faced symbols for particular images under a representation.  Both kinds
of symbols may appear in a single expression, in which case a map defined by
pointwise multiplication is described as in $[\gamma_\mu \bbox\gamma](a) =
\gamma_\mu [\bbox\gamma(a)]$, $a \in {\cal C \it l}(M)$.)
With this definition we obtain the familiar relationship
\begin{equation}
\{\gamma_\mu, \gamma_\nu\} =  2\, {\rm g}_{\mu\nu} \bbox{1}.
\label{eq:repcomps}
\end{equation}
The Dirac spinor bundle $S(M)$ is a vector bundle carrying such a
representation $\bbox\gamma$, i.e., there is a chart for $S(M)$ such that the
Clifford action of ${\cal C \it l}(M)$ on $S(M)$ is given by multiplication of
the $\gamma$-matrices with column spinors.  If ${\cal C \it l}(M)$ transforms
under a product of tensor representations of the orthogonal group and the
Clifford action is covariant under this transformation, then $S(M)$ transforms
under a spin representation of the orthogonal group.

Except for regions that contain zeroes of ${h}$ it is straightforward to
generalize these ideas to our signature changing spacetime $M$.
The crucial question is how to link the spinor bundles across hypersurfaces of
signature change.
In the following exposition we will use the fact that the Clifford bundles are
linked and lift this link to the spinor bundles.  Specifically we
will consider an algebra fibration which coincides with the Clifford bundles
where the metric is non-degenerate and representations of this fibration which
are continuous
across a hypersurface of signature change.
A detailed study of such
representations suggests
certain additional conditions which are
sufficient to ensure the invariance of the resulting structure under
appropriate changes of representations and/or coordinates.
Since the group of transition functions is different for different signature
we will adopt the term ``fibration'' for ${\cal C \it l}(M)$ and $S(M)$
instead of ``bundle'', but we will still refer to this object as the
``Clifford'' and ``spinor'' fibration, although we use these expressions in a
non-traditional context.

The following example will illustrate some of the key issues we have to face.

\subsection{An example in two dimensions}

In $n=2$ dimensions we consider $M={\Bbb R}^2$ with coordinates $(t,x)$ and
metric
\begin{equation}
{\rm g} = {h(t)}\, dt \otimes dt + dx \otimes dx,
\label{eq:metric}
\end{equation}
i.e., $\hat{\rm g} = 1$.
Then the following $\gamma$-matrices
\begin{equation}
\gamma_x=\pmatrix{1&\phantom{-}0\cr 0&-1\cr},\quad
\gamma_t=\pmatrix{0&1\cr {h(t)} &0\cr},
\label{eq:gamma}
\end{equation}
define a continuous representation $\bbox\gamma$ on all of $M$ which is
faithful and irreducible for ${h(t)} \not= 0$.
Note that $\gamma_t$ is necessarily degenerate at zeroes of ${h}$, where the
matrix algebra generated by these matrices actually reduces to
upper-triangular matrices.  Therefore, this representation is neither faithful
nor irreducible at metric degeneracies.  This behavior is generic because of
an incompatibility
of representations of degenerate and non-degenerate Clifford algebras: Since
the Clifford algebra is no longer semi-simple for ${h} = 0$, the dimension
of an irreducible representation is smaller by a factor of two.  The
irreducible representation of the degenerate algebra is in fact just an
irreducible representation of its non-degenerate ``spatial'' part, i.e., the
part corresponding to the ``spatial'' $\gamma_x$.
For the representation to remain faithful it would have to double its
dimension in order to accommodate the whole nilpotent ideal generated by the
degenerate direction.  (Note that half of the algebra, namely the ideal
generated by $\gamma_t$, is nilpotent of order~2 at the degeneracy.)

\subsection{The general case}

For a precise description of the behavior of a representation around a metric
degeneracy we examine the behavior of $\bbox\gamma(\partial_\mu)|_p$ as $p$
approaches a hypersurface ${H} = \{t =t_0\}$, where ${h}(t_0) = 0$.
\begin{observation}
If a continuous local representation $\bbox\gamma$ satisfies
Eq.~(\ref{eq:rep}) on an open set $U \subset M$ intersecting ${H}$ and is
faithful and irreducible on $U \backslash {H}$, then $\bbox\gamma$ is a
faithful representation of the ``spatial subalgebra'' ${\cal C \it l}^{sp}(M)$
generated by $\{\partial_i\}_{i = 1,\ldots,n-1}$ on all of $U$.  Furthermore,
${\cal C \it l}^{sp}(M)$ contains central orthogonal idempotents $P_\pm$ which
effect a Pierce decomposition of ${\cal C \it l}(M)$ and a corresponding
decomposition of $\bbox\gamma$.  Given a particular form of
$\bbox\gamma(P_\pm)$, this decomposition is reflected in a block structure of
the matrix representation.
\end{observation}
 From the previous example we infer that $\bbox\gamma(\partial_t)|_p$ becomes
degenerate as $p \to {H}$.  For the other coordinate vector fields this is
not the case, since $\bbox\gamma(\partial_i)^2 = \bbox 1$ everywhere in $U$.
This corresponds to the fact that the algebra generated by $\{\partial_i\}_{i
= 1,\ldots,n-1}$, which we call the ``spatial subalgebra'' ${\cal C \it
l}^{sp}(M)$ remains non-degenerate on ${H}$, whence $\bbox\gamma$ restricted
to ${\cal C \it l}^{sp}(M)$ remains a faithful representation.
Therefore, this spatial subalgebra does not ``notice'' the metric degeneracy
and will provide the link that constrains the behavior of $\gamma_t$ as we
pass through ${H}$.
${\cal C \it l}^{sp}(M)$ contains central orthogonal idempotents
\begin{equation}
P_\pm := {1\over 2}(1 \pm z),
\end{equation}
where $z$ is the normalized dual of the volume element of ${H}$, whence $z^2 =
1$, $P_\pm^2 = P_\pm$, and $P_\pm P_\mp = 0$.  For example, $z = \partial_x$
for the metric given by Eq.~(\ref{eq:metric}), whereas $z = i \,
(\det{\hat{\rm g}})^{-{1\over 2}} \partial_1 \wedge \partial_2 \wedge
\partial_3$ in four dimensions with metric given by Eq.~(\ref{eq:metric4}).
The idempotents or projectors $P_\pm$ split ${\cal C \it l}^{sp}(M)$ into a
direct sum of simple components
\begin{equation}
{\cal C \it l}^{sp}(M) = {\cal C \it l}_+(M) \oplus {\cal C \it l}_-(M),
\end{equation}
where
\begin{equation}
{\cal C \it l}_\pm(M) := P_\pm {\cal C \it l}(M) P_\pm.
\end{equation}
Therefore
$\bbox\gamma$ induces inequivalent representations
\begin{equation}
\bbox\gamma_\pm := \gamma_\pm \bbox\gamma \gamma_\pm
\label{eq:rep+-}
\end{equation}
of ${\cal C \it l}^{sp}(M)$, where
\begin{equation}
\gamma_\pm := \bbox\gamma(P_\pm).
\end{equation}
So we get the following Pierce decomposition with respect to the idempotents
$P_\pm$:
\begin{eqnarray}
{\cal C \it l}(M)
&=& (P_+ + P_-) \, {\cal C \it l}(M) \, (P_+ + P_-)\nonumber\\
&=& P_+ {\cal C \it l}(M) P_+ \oplus P_+ {\cal C \it l}(M) P_- \oplus
P_- {\cal C \it l}(M) P_+ \oplus P_- {\cal C \it l}(M) P_-
\nonumber\\[\smallskipamount]
&=&
{\cal C \it l}_+(M) \oplus P_+ {\cal C \it l}(M) P_- \oplus
P_- {\cal C \it l}(M) P_+ \oplus {\cal C \it l}_-(M),\label{eq:algebra-Pierce}
\end{eqnarray}
which translates into representations
\begin{eqnarray}
\bbox\gamma &=&
\gamma_+\bbox\gamma\gamma_+ + \gamma_+\bbox\gamma\gamma_- +
\gamma_-\bbox\gamma\gamma_+ + \gamma_-\bbox\gamma\gamma_-
\nonumber\\[\smallskipamount]
&=& \bbox\gamma_+ + \gamma_+\bbox\gamma\gamma_- +
\gamma_-\bbox\gamma\gamma_+ + \bbox\gamma_-
\end{eqnarray}
Since ${\cal C \it l}^{sp}(M)$ commutes with $P_\pm$ and
\begin{equation}
{\cal C \it l}(M)
= {\cal C \it l}^{sp}(M) \oplus {\cal C \it l}^{sp}(M) \partial_t
= {\cal C \it l}^{sp}(M) \oplus \partial_t {\cal C \it l}^{sp}(M),
\end{equation}
the cross terms in Eq.~(\ref{eq:algebra-Pierce}) come from $\partial_t$:
\begin{eqnarray}
P_\pm {\cal C \it l}(M) P_\mp
&=& P_\pm ({\cal C \it l}^{sp}(M)\, \partial_t) P_\mp
= {\cal C \it l}_\pm(M) \, \partial_t
\nonumber\\[\smallskipamount]
&=& P_\pm (\partial_t\, {\cal C \it l}^{sp}(M)) P_\mp
= \partial_t \, {\cal C \it l}_\mp(M).\label{eq:algebra-off-diagonal}
\end{eqnarray}
(Note that $P_\pm \partial_t = \partial_t P_\mp$ and $P_\pm {\cal C \it
l}^{sp}(M) P_\mp = 0$.)  This can also be seen from the decomposition of
$\gamma_t$:
\begin{equation}
\gamma_t = \gamma_+ \gamma_t \gamma_- + \gamma_- \gamma_t \gamma_+
\end{equation}
If $\gamma_\pm$ take the form
\begin{equation}
\gamma_+ = \pmatrix{ \bbox 1 &\bbox 0 \cr \bbox 0  & \bbox 0\cr},
\qquad
\gamma_- = \pmatrix{ \bbox 0 &\bbox 0 \cr \bbox 0  & \bbox 1\cr},
\label{eq:block-proj}
\end{equation}
in terms of $2^{m-1} \times 2^{m-1}$ unit and zero matrices, which can always
be achieved by an equivalence transformation pointwise on $U$ (even on ${H}$),
then the Pierce decomposition is reflected in a block structure of the matrix
representation $\bbox\gamma({\cal C \it l}(M))$.  In particular, the induced
representations $\bbox\gamma_\pm$ only have one non-zero block, namely in the
upper left (lower right) corner.
Denoting the non-zero blocks of the corresponding matrices by overlined
symbols, for example,
\begin{equation}
\bbox\gamma_+({\cal C \it l}_+(M)) =
\pmatrix{
\overline{\bbox\gamma_+({\cal C \it l}_+(M))}&\bbox0\cr
\bbox0&\bbox0\cr}\,,
\end{equation}
we have the following block structure of $\bbox\gamma({\cal C \it l}(M))$:
\begin{eqnarray}
\bbox\gamma({\cal C \it l}(M)) &=&
\pmatrix{
\overline{\bbox\gamma_+({\cal C \it l}_+(M))}&
\overline{\gamma_+ \gamma_t \bbox\gamma_-({\cal C \it l}_-(M))} \cr
\noalign{\smallskip}
\overline{\gamma_- \gamma_t \bbox\gamma_+({\cal C \it l}_+(M))}
&  \overline{\bbox\gamma_-({\cal C \it l}_-(M))}\cr}
\nonumber\\[\medskipamount]
&=&
\pmatrix{
\overline{\bbox\gamma_+({\cal C \it l}_+(M))}
& \overline{\bbox\gamma_+({\cal C \it l}_+(M)) \gamma_t \gamma_-} \cr
\noalign{\smallskip}
\overline{\bbox\gamma_-({\cal C \it l}_-(M)) \gamma_t \gamma_+}
& \overline{\bbox\gamma_-({\cal C \it l}_-(M))}\cr}.
\label{eq:algebra-blocks}
\end{eqnarray}
(To arrive at this equation apply $\bbox\gamma$ to
Eq.~(\ref{eq:algebra-Pierce}) using
Eqs.~(\ref{eq:rep+-},\ref{eq:algebra-off-diagonal},\ref{eq:block-proj}) and
inserting projectors $P_\pm$ when appropriate.)
This block structure helps us to understand what happens to a representation
when we cross ${H}$.  The blocks on the diagonal make up the spatial
subalgebra and do not contain $\gamma_t$.  Therefore, these blocks remain
non-degenerate throughout $U$.  The off-diagonal blocks show that $\gamma_t$
intertwines $\bbox\gamma_+$ and $\bbox\gamma_-$.
\begin{observation}
The inequivalent faithful representations $\bbox\gamma_\pm$ of ${\cal C \it
l}^{sp}(M)$ have equivalent restrictions $\bbox\gamma^+_\pm$ to the even
subalgebra ${\cal C \it l}^+(M) \subset {\cal C \it l}^{sp}(M)$.  Furthermore,
the restrictions $\bbox\gamma^+_\pm$ are intertwined by $\gamma_t$, which
implies that for any $p \in {H}$, one of the off-diagonal blocks of
$\gamma_t|_p$ in the previously discussed block structure vanishes and the
other either vanishes or is regular.  (The diagonal blocks are trivially
zero.)
\end{observation}
Even though the representations $\bbox\gamma_\pm$ vanish on one of the simple
components, $\bbox\gamma_\pm({\cal C \it l}_\mp(M)) = 0$, they are equivalent
when restricted to the even part ${\cal C \it l}^+(M)$ of ${\cal C \it
l}^{sp}(M)$, which is a simple algebra isomorphic to ${\cal C \it l}_\pm(M)$.
Applying $\gamma_i \gamma_t = -\gamma_t \gamma_i$ twice, we have $\gamma_i
\gamma_j \gamma_t = + \gamma_t \gamma_i \gamma_j$, which implies that the
restrictions $\bbox\gamma_\pm^+$ of $\bbox\gamma_\pm$ to ${\cal C \it l}^+(M)$
are intertwined by $\gamma_t$:
\begin{equation}
\bbox\gamma_\pm^+ \gamma_t = \gamma_t \bbox\gamma_\mp^+.
\label{eq:intertwines}
\end{equation}
In the block structure (\ref{eq:algebra-blocks}) the non-zero blocks
$\overline{\bbox\gamma_\pm^+({\cal C \it l}^+(M))}$ induce irreducible
representations $\overline{\bbox\gamma_\pm^+}:\pi^{-1}(U) \cap {\cal C \it
l}^+(M) \to M_{k\over 2}({Bbb C}) \times U$.
Since an intertwiner of two irreducible representations is determined up to a
scale, with the intertwiner being non-singular unless the scale is zero, we
see from the non-zero blocks associated with Eq.~(\ref{eq:intertwines}) that
the two blocks of $\gamma_t = \gamma_+ \gamma_t \gamma_- + \gamma_- \gamma_t
\gamma_+$ are determined by Eq.~(\ref{eq:intertwines}) up to a scale.  Since
$(\gamma_\pm \gamma_t \gamma_\mp) (\gamma_\mp \gamma_t \gamma_\pm) = {h(t)}
\gamma_\pm$, in fact, only a relative scale remains undetermined.
Therefore at least one entire block of $\gamma_t$ has to vanish for ${h(t)}
\to 0$, so that we are left with a block triangular or block diagonal matrix
algebra on ${H}$.

Even though we may not be able to achieve this block structure on all of $U$
at the same time, this argument still shows that $\gamma_t$ is determined up
to a relative scale between $\gamma_+ \gamma_t \gamma_-$ and $\gamma_-
\gamma_t \gamma_+$ and that $\bbox\gamma({\cal C \it l}(M))$ is isomorphic to
a block triangular or block diagonal matrix algebra at each point of ${H}$.
\begin{observation}
Two continuous local representations $\bbox\gamma_{(r)}$, $r = 1$, $2$,
satisfying Eq.~(\ref{eq:rep}) on an open set $U \subset M$ intersecting
${H}$ and faithful irreducible on $U \backslash {H}$, are equivalent if
and only if the block structures of $\bbox\gamma_{(r)}(\partial_t)|_{{H}}$
agree.  Furthermore, the intertwiner is guaranteed to be continuous across
${H}$ if one block of $\bbox\gamma_{(r)}(\partial_t)$ stays regular.
\end{observation}
Given two overlapping local representations, we can use the same decomposition
to show that it is a necessary condition that $\gamma_t$ has the same behavior
on ${H}$ for both representations if they are related by a non-singular
intertwiner.
Conversely, if the behavior of $\gamma_t$ is different for two local
representations, the intertwiner necessarily becomes singular on ${H}$.
Not only the agreement in block structure but its particular form on ${H}$
is of importance.  If both blocks of $\gamma_t$ vanish, i.e., $\gamma_t$
vanishes entirely for both overlapping local representations, their
intertwiner may be discontinuous.
If on the other hand only one block of $\gamma_t$ vanishes then the
intertwiner inherits the smoothness properties of the local representations,
in particular it is at least continuous.  In this case the non-zero block of
$\gamma_t$ serves as a link across ${H}$ and no additional requirement of
continuity of the intertwiner is needed to ensure that the gluing together of
local representations is well-defined.  Of course, the transition functions
can be restricted to lie in the appropriate spin groups away from ${H}$,
which requires the transition functions on ${H}$ to continuously connect
both spin groups.

\subsection{Criteria for a spinor fibration}

It can be shown
that if $\bbox\gamma$ is assumed to be not only $C^1$ away from ${H}$
(this is required in order to define a spin connection as we will see in
section \ref{sec:spinor conn}) but also to have bounded partial derivatives
on any bounded set, then exactly one of $\gamma_\pm \gamma_t
\gamma_\mp$ vanishes on all of ${H}$ and the other one does not.
Therefore, a simple smoothness assumption gains the desired control over the
block structure.  Since the minor technical difference between requiring
bounded partial derivatives on bounded sets and $C^1$, namely that the partial
derivatives have limits on ${H}$, does not affect the continuity structure
of the spinor fibration in question, we will use the more intuitive condition
of continuous differentiability.
Allowing the partial derivatives of $\bbox\gamma$ to be locally unbounded
relinquishes any control over the block structure, e.g., in the two
dimensional example:
\begin{equation}
\gamma_x=\pmatrix{1&\phantom{-}0\cr 0&-1\cr},\quad
\gamma_t=\left\{\matrix{
\pmatrix{0&|{h(t)}|^{1\over 2} + x^2\cr
{h(t)}[|{h(t)}|^{1\over 2} + x^2]^{-1} &0\cr} & \text{for\ } x > 0\cr
\noalign{\smallskip}
\pmatrix{0&|{h(t)}|^{1\over 2}\cr
{h(t)}|{h(t)}|^{-{1\over 2}} &0\cr} & \text{for\ } x \leq 0\cr}\right..
\label{eq:strange}
\end{equation}
Piecing $\bbox\gamma$'s like this one together we can get any behavior of
$\gamma_t$ on ${H}$ we (don't) like.

These observations lead us to a set of criteria for local representations
which ensure that they are related by $C^1$ equivalence transformations:
\begin{itemize}
\item[(i)]
$\bbox\gamma$ is $C^1$ satisfying Eq.~(\ref{eq:rep}).
\item[(ii)]
$\bbox\gamma$ is faithful irreducible for ${h(t)} \not= 0$.
\item[(iii)]
$\gamma_-\gamma_t \to 0$ for ${h(t)} \to 0$.
\end{itemize}
(Of course, the $\gamma$-matrices given by Eq.~(\ref{eq:gamma}) satisfy these
criteria.)
Condition (iii) singles out one class of representations with a certain
behavior for ${h(t)} \to 0$.  Equally well, one could require
\begin{itemize}
\item[(iii${}'$)]
$\gamma_+\gamma_t \to 0$ for ${h(t)} \to 0$.
\end{itemize}
or even a mixture of both, fixing the behavior of the representation for each
hypersurface of metric degeneracy separately.  In this paper we focus on the
issues arising from just one zero of ${h}$.
In this case (iii${}'$) is obtained from (iii) under a spatial inversion.

\subsection{A possible generalization}

We can relax the assumption of a metric of the form Eq.~(\ref{eq:metric4}) if
we assume the existence of a local frame of non-zero vector fields $\{X_\mu\}$
on any open set intersecting a hypersurface ${H}$ of signature change, such
that
$X_i \in T({H})$ satisfies ${\rm g}(X_i,X_j)|_{H} = \delta_{ij}$ and ${\rm
g}(X_0,X_\mu)|_{H} = 0$ and construct $\bbox\gamma(X_\mu)$ instead of
$\bbox\gamma(\partial_\mu)$.  The spatial subalgebra ${\cal C \it l}^{sp}(M)$
of ${\cal C \it l}(M)$ generated by $\{X_i\}$ coincides with the appropriate
extension of ${\cal C \it l}({H}) = \mathop{\bigcup}\limits_{p \in {H}} {\cal
C \it l}(T_p{H},{H}^*{\rm g}_p)$, which is really the only intrinsic structure
in the vicinity of ${H}$.  It is essential that the pullback metric ${H}^*{\rm
g}_p$ be non-degenerate.  It is then straightforward to retrace the steps we
followed above and come to the same conclusions.  Of course, the existence of
a global fibration $S(M)$ will depend on the topology of $M$ and possibly on
the topology of hypersurfaces of metric degeneracy.

\section{The spinor covariant derivative}\label{sec:spinor cov}

Having defined a spinor fibration $S(M)$ we have a notion of continuity of a
spinor field.  Namely, a spinor field is continuous if its component sections
are continuous with respect to a bundle chart.  In other words, given a set of
$\gamma$-matrices satisfying appropriate conditions, a continuous spinor field
is given by a column of continuous functions on which these $\gamma$-matrices
act.

In order to write down a Dirac equation on $M$, we need a notion of covariant
differentiation of a spinor field.  However, given a linear connection on
$M$, the spinor connection is not uniquely determined unless it is
also required to be compatible with both a choice of spinor metric and a
notion of charge conjugation.
Furthermore, the traditional construction of a
spinor connection relies on the existence of a non-degenerate metric.  In the
following we discuss these separate aspects in regions where the metric is
manifestly non-degenerate.  In section \ref{sec:Dirac} the interrelation
between these different aspects will be examined in the vicinity of a
hypersurface of signature change.

Authors of other literature on this subject usually work in orthonormal frames
(see for example \cite{Penrose and Rindler}) with the notable exception of an
early review \cite{Bade and Jehle} which also contains references to most of
the original work and notes the scaling freedom in the spinor metric discussed
below.

\subsection{The spinor metric}

In order to discuss the Dirac equation below we introduce the notion of a
spinor metric.  In particular, we adopt a hermitian symmetric spin invariant
bilinear form on Dirac spinors
\begin{equation}
\vbox{\halign{$\hfil#\hfil$&${}#{}$&$#\hfil$\cr
S(M) \times S(M)&\to&{\frak F}(M)\cr\noalign{\smallskip}
\Psi,\Xi&\mapsto&{(\Psi,\Xi)} = \Psi^\dagger C\, \Xi,\cr}}
\end{equation}
where ${\frak F}(M)$ denotes the space of functions on $M$ and $C$ is chosen
to satisfy
\begin{eqnarray}
C &=& C^\dagger,\label{eq:Ca}\\
C \gamma_\mu &=& - \gamma_\mu^\dagger C \label{eq:Cb}
\end{eqnarray}
on $M$.  The familiar Dirac adjoint is then given by
\begin{equation}
\bar\Psi = \Psi^\dagger C.
\end{equation}
For our example Eq.~(\ref{eq:gamma})
\begin{equation}
C_1 = \pmatrix{ 0& -i\cr i &0 \cr}
\end{equation}
satisfies Eqs.~(\ref{eq:Ca}) and (\ref{eq:Cb}).
However, the spinor metric $C$ is only determined up to a real scalar at each
point of the manifold.  Therefore $C_f = f C_1$ could equally well be chosen
as spinor metric, where $f = f^{*}\in {\frak F}(M)$.  Usually the spinor
metric is required to be smooth and non-degenerate, which restricts $f$ to be
smooth and non-zero.
This is one of the reasons why the choice of spinor metric does not usually
appear in the standard discussion of the Dirac equation.
The scaling function $f$ is normalized to make
the equation simple, i.e., $C$ is chosen to be constant for constant
$\gamma$-matrices.  (Note that the $\gamma$-matrices cannot be constant
across a hypersurface of signature change.)
The behavior of $f$ where the spacetime metric is degenerate must be
postulated separately, and it can not {\it a priori} be ruled out that $f$ may
be zero or singular there.

\subsection{Charge conjugation}

Charge conjugation can be defined as a map
\begin{equation}
\vbox{\halign{$\hfil#\hfil$&${}#{}$&$#\hfil$\cr
S(M) &\to&S(M)\cr\noalign{\smallskip}
\Psi&\mapsto&\Psi^c := B^{*} \Psi^{*},\cr}}
\end{equation}
where $B$ satisfies
\begin{eqnarray}
B \gamma_\mu &=& \gamma_\mu^* B \label{eq:Ba},\\
\quad B^*B &=& \pm {\bf 1} = \beta\, {\bf 1}.\label{eq:Bb}
\end{eqnarray}
These conditions determine $B$ up to a phase which may vary over $M$.  The
sign in the second condition depends on the signature.
$\beta = +1$ if there exists a real representation $\beta = -1$ otherwise.
Defining the index $\nu$ of a metric to be the (signed) difference of the
number of
positive and negative eigenvalues of the metric we note that
\begin{equation}
\beta
= \left\{
{+1 \text{\ for\ } \nu \equiv 0,2 \text{\ mod\ } 8 \atop
-1 \text{\ for\ } \nu \equiv 4,6 \text{\ mod\ } 8} \right.
\label{beta}
\end{equation}
Therefore, $\beta$ changes sign and $B$ is necessarily discontinuous if the
signature changes from $(-+++)$, i.e., $\nu = 2$, to $(++++)$, i.e., $\nu =
4$, in four dimensions, while for the change of signature $(-+) \to (++)$ in
two dimensions $\beta = 1$ in both regions.  Since $\beta$ also determines
the periodicity of the charge conjugation operation, namely
\begin{equation}
(\Psi^c)^c = \beta \, \Psi,
\end{equation}
continuity of a spinor is only compatible with continuity of its charge
conjugate if $\beta$ is the same in Euclidean and Lorentzian regions.
(This observation warrants an investigation of alternative
spinor metrics and notions of charge conjugation for the opposite metrics,
i.e., signature changing from $(+---)$ to $(----)$, in four dimensions.  The
reader is invited to pursue these technical aspects which lie outside the main
thrust of this article.  Note that the standard definitions for opposite
Lorentzian metrics differ by signs in Eqs.~(\ref{eq:Cb}) and (\ref{eq:Ba}).
For completeness, one may also consider the inclusion of spinors with
Grassmann-valued components or even non-standard versions of
Eqs.~(\ref{eq:Cb}) and (\ref{eq:Ba}).)

For our 2-dimensional example, we may take
\begin{equation}
B = e^{i\theta} {\bf 1},
\end{equation}
where $\theta = \theta^* \in {\frak F}(M)$.

\subsection{The spinor connection}\label{sec:spinor conn}

Given a spinor metric the spinor covariant derivative $S_\mu$ with respect to
a vectorfield $\partial_\mu$ is given by
\begin{equation}
S_\mu = \partial_\mu + \Sigma_\mu,
\label{eq:S}\end{equation}
where the spinor connection $\Sigma_\mu$ has to be determined such that the
axioms for a spinor covariant derivative are satisfied:
\begin{eqnarray}
S_\mu (a^\nu \gamma_\nu \Psi)
&=& (\nabla_\mu a^\nu) \gamma_\nu \Psi + a^\nu \gamma_\nu (S_\mu\Psi),
\label{eq:sp connection}\\[\smallskipamount]
\partial_\mu {(\Psi,\Xi)}
&=& {( S_\mu\Psi,\Xi)} + {(\Psi,{S_\mu\Xi})},
\label{eq:sp metricity}\\[\smallskipamount]
S_\mu(\Psi^c) &=& (S_\mu\Psi)^c.\label{eq:sp charge}
\end{eqnarray}
$\nabla_\mu a^\nu := {a^\nu}_{;\mu} := \partial_\mu a^\nu +
{\Gamma^\nu}_{\mu\rho} \, a^\rho$ denotes the components of the covariant
derivative of the vector field given by $a^\nu$, where $\Gamma_{\rho\mu\nu}$
are the spacetime connection coefficients,
i.e., for the Levi-Civita connection $\Gamma_{\rho\mu\nu} = {1\over 2}
(\partial_\mu {\rm g}_{\nu\rho} + \partial_\nu {\rm g}_{\mu\rho}
- \partial_\rho {\rm g}_{\mu\nu})$.
These axioms ensure compatibility of covariant differentiation of tensors and
spinors, Eq.~(\ref{eq:sp connection}), and compatibility of the spinor
covariant derivative with the spinor metric and charge conjugation,
Eqs.~(\ref{eq:sp metricity}) and (\ref{eq:sp charge}).
Using the defining properties
Eqs.~(\ref{eq:S},\ref{eq:Ca},\ref{eq:Cb},\ref{eq:Ba},\ref{eq:Bb}) in
Eqs.~(\ref{eq:sp connection}-\ref{eq:sp charge}) we get the following
conditions:
\begin{eqnarray}
\partial_\mu \gamma_\nu - {\Gamma^\rho}_{\mu\nu} \gamma_\rho
&=& [\gamma_\nu,\Sigma_\mu] = \gamma_\nu \Sigma_\mu - \Sigma_\mu\gamma_\nu
\label{eq:Si connection}\\[\smallskipamount]
C^{-1}\partial_\mu C
&=& \Sigma_\mu + C^{-1} \Sigma_\mu^\dagger C,
\label{eq:Si metricity}\\[\smallskipamount]
B^{-1}\partial_\mu B
&=& \Sigma_\mu - B^{-1} \Sigma_\mu^* B.\label{eq:Si charge}
\end{eqnarray}

In order to give an explicit expression for $\Sigma_\mu$ we expand it in
a basis of the Clifford algebra:
\begin{equation}
\Sigma_\mu = \sum_I \sigma_{\mu\,I} \gamma^I,
\end{equation}
where the sum is taken over the set of ordered indices $\{(i_1, \ldots, i_p) :
1 \leq i_1 < \cdots < i_p \leq n, 0 \leq p \leq n - 1\}$, with $n = \dim M$,
where also
$\gamma^{(i_1 \! \ldots i_p)} = \gamma^{i_1} \ldots \gamma^{i_p}$ and
$\gamma^{{\emptyset}} = {\bf 1}$ are understood.
(Note that the superscript is the empty set $\emptyset$ {\em not} $0$ in the
last equation.)
In particular $\{\gamma^I \}$ is a
basis for the Clifford algebra in the representation $\bbox\gamma$.

We first solve for the components of $\Sigma_\mu$ using Eq.~(\ref{eq:Si
connection}):
\begin{equation}
[\gamma_\nu,\Sigma_\mu] = [\gamma_\nu,\sum_I \sigma_{\mu\,I} \gamma^I] =
2 \sum_{\hbox to 0pt{\hss
$\scriptscriptstyle {\nu \not\in I \atop |I| \text{\ odd}}$
\hss}}
\sigma_{\mu\,I} \gamma_\nu \gamma^I
+ 2 \sum_{\hbox to 0pt{\hss
$\scriptscriptstyle {\nu \in I \atop |I| \text{\ even}}$
\hss}}
\sigma_{\mu\,I} \gamma_\nu \gamma^I
\end{equation}
where $|I|$ denotes the length of the multi index.  Thus all but the scalar
part of $\Sigma_\mu$ is determined:
\begin{equation}
\sigma_{\mu\,I}
= {1\over 2^{N+1}}
{\mathop{\rm tr\;}\nolimits}[
\gamma_{I^r} \gamma^\nu (\partial_\mu \gamma_\nu -
{\Gamma^\rho}_{\mu\nu} \gamma_\rho)
]
\qquad (\text{no sum over\ }\nu),
\label{eq:si connection}
\end{equation}
where for given $I$ one may choose any $\nu$ such that for $|I|$ even  $\nu
\in I$ while for $|I|$ odd  $\nu \not\in I$.  ($I^r$ denotes indices in
reversed order, $N = 2^{n \over 2}$.)  For example, to calculate
$\sigma_{\mu\,(0,1,2,3)}$ in four dimensions we may take any $\nu \in
\{0,1,2,3\}$, the result is guaranteed to be the same.

We solve for the scalar part of $\Sigma_\mu$ using Eqs.~(\ref{eq:Si
metricity}) and (\ref{eq:Si charge}):
\begin{equation}
\sigma_{\mu \, {{\emptyset}}}
= {1\over 2^{N+1}}
{\mathop{\rm tr\;}\nolimits}(C^{-1}\partial_\mu C + B^{-1}\partial_\mu B).
\label{eq:si scalar}
\end{equation}
Thus $\Sigma_\mu$ is completely determined.
Eq.~(\ref{eq:si scalar}) is derived from the general conditions arising from
Eqs.~(\ref{eq:Si metricity}) and (\ref{eq:Si charge}):
\begin{eqnarray}
{\mathop{\rm Re\;}\nolimits} \sigma_{\mu\,I}
&=& {1\over 2^{N+1}}
{\mathop{\rm tr\;}\nolimits}(\gamma_{I^r} C^{-1}\partial_\mu C)
\qquad
(|I| \text{\ even}),
\\[\smallskipamount]
{\mathop{\rm Im\;}\nolimits} \sigma_{\mu\,I}
&=& {1\over 2^{N+1}i}
{\mathop{\rm tr\;}\nolimits}(\gamma_{I^r} B^{-1}\partial_\mu B).
\end{eqnarray}
Again these expressions are guaranteed to be real and compatible with
Eq.~(\ref{eq:si connection}).  In some instances it may
actually be more convenient to use these latter relationships to solve for
various components of $\Sigma_\mu$.

Applying Eqs.~(\ref{eq:si connection},\ref{eq:si scalar}) to
Eqs.~(\ref{eq:gamma},\ref{eq:Ca},\ref{eq:Cb},\ref{eq:Ba},\ref{eq:Bb}) we
obtain for the spinor connection for our 2-dimensional example
\begin{eqnarray}
\Sigma_x &=&
{1\over 2} \, f^{-1} \partial_x f + {1\over 2} \, i \, \partial_x \theta
\nonumber\\[\smallskipamount]
\Sigma_t &=&
{1\over 2} \, f^{-1} \partial_t f + {1\over 2} \, i \, \partial_t \theta +
{1\over 4} \, {h}^{-1} \partial_t {h} \, \gamma_x
\label{eq:Sigma}
\end{eqnarray}

In the case of a local orthonormal frame $\{X_a\}$ with constant
$\gamma$-matrices and constant matrices $C$ and $B$, the familiar solution for
$\Sigma_a$ is purely a bivector
\begin{equation}
\Sigma_a = {1\over 4} \,\omega_{abc} \gamma^b\gamma^c,
\end{equation}
where $\omega_{abc} = {\rm g}(X_b,\nabla^{\strut}_{X_a}X_c)$ are the
connection coefficients.  (Note that the metric compatibility of the
connection implies $\omega_{abc} = - \omega_{acb}$.)

\section{The massless Dirac equation in two dimensions}\label{sec:Dirac}

With the definition (\ref{eq:S}) of the spinor covariant derivative
the massless
Dirac equation in arbitrary dimensions takes the form
\begin{equation}
{{{S\mkern -2mu}\llap{/}}\mkern 2mu}
\Psi \equiv \gamma^\mu S_\mu \Psi = 0.
\label{eq:Dirac}
\end{equation}
In two dimensions for the spinor connection (\ref{eq:Sigma}) we
obtain a family of equations depending on the two real functions $f$ and
$\theta$:
\begin{equation}
[\gamma^\mu (\partial_\mu + {1\over 2} \, f^{-1} \partial_\mu f +
{1\over 2} \, i \, \partial_\mu \theta)
+ \gamma^t {1\over 4} \, {h}^{-1} \partial_t {h} \, \gamma_x] \Psi = 0.
\label{eq:Dirac 2d}
\end{equation}

\subsection{Solution for the massless Dirac equation in two dimensions}

We solve this equation for regions where it is regular.
It is easy to check that Eq.~(\ref{eq:Dirac 2d}) is equivalent to
\begin{equation}
\gamma^\mu
[(f^{-{1\over 2}}e^{-{1\over 2}i\,\theta}  D^{-1}) \,
\partial_\mu \,
(f^{1\over 2}e^{{1\over 2}i\,\theta} D)] \Psi = 0,
\end{equation}
where the matrix $D$ must satisfy
\begin{eqnarray}
\partial_t D &=&
D {1\over 4} \, {h}^{-1} \partial_t {h} \, \gamma_x,
\\[\smallskipamount]
\partial_x D &=& 0.
\end{eqnarray}
Thus, up to an unimportant constant factor,
\begin{equation}
D = {1 \over 2}
[|{h}|^{1 \over 4}(1 + \gamma_x) + |{h}|^{-{1 \over 4}}(1 - \gamma_x)]
= \pmatrix{ |{h}|^{1 \over 4} & 0 \cr  0 & |{h}|^{-{1 \over 4}} \cr},
\end{equation}
with
\begin{equation}
D^{-1} = {1 \over 2}
[|{h}|^{-{1 \over 4}}(1 + \gamma_x) + |{h}|^{1 \over 4}(1 - \gamma_x)]
= \pmatrix{ |{h}|^{-{1 \over 4}} & 0 \cr  0 & |{h}|^{1 \over 4} \cr}.
\end{equation}
The plane wave ansatz
\begin{equation}
\Psi = (f^{-{1\over 2}}e^{-{1\over 2}i\,\theta}  D^{-1}) \psi_0 \,
e^{-i (k_\tau \tau - k_x x)},
\end{equation}
where $\tau = {\displaystyle\int} \sqrt{|{h(t)}|} \, dt$, leads to
\begin{equation}
(- \gamma^t \sqrt{|{h}|} k_\tau + \gamma^x k_x) D^{-1} \psi_0 = 0.
\end{equation}
For non-trivial solutions we need
\begin{equation}
\det(- \gamma^t \sqrt{|{h}|} k_\tau + \gamma^x k_x)
= - k_x^2 - {h}^{-1}|{h}| k_\tau^2
= 0,
\end{equation}
which gives the dispersion relation
\begin{equation}
k_\tau = \left\{\matrix{
\pm k_x & \text{for\ } {h} < 0\cr \pm i \, k_x & \text{for\ } {h} >
0\cr}\right.,
\end{equation}
and corresponding solutions for $\psi_0$
\begin{eqnarray}
\psi_0 &=& \pmatrix{1 \cr \mp 1 \cr} \qquad
\text{for\ } {h} < 0, k_\tau = \pm k_x,
\\[\smallskipamount]
\psi_0 &=& \pmatrix{1 \cr \mp i \cr}
\qquad \text{for\ } {h} > 0, k_\tau = \pm i \, k_x.
\end{eqnarray}
Thus the general solutions
for regions where ${h} \not= 0$ and $f \not= 0$ are:
\begin{eqnarray}
\Psi_L = (f^{-{1\over 2}}e^{-{1\over 2}i\,\theta}  D^{-1})
&\displaystyle \sum_{k > 0}&
\left[(a_k^+ e^{i k x}\pmatrix{1 \cr -1 \cr}
+ a_k^- e^{-i k x}\pmatrix{1 \cr 1 \cr}) \, e^{-i k \tau}\right.
\nonumber\\[\smallskipamount]
&&\left.{}+
(b_k^+ e^{i k x}\pmatrix{1 \cr 1 \cr}
+ b_k^- e^{-i k x}\pmatrix{1 \cr -1 \cr}) \, e^{i k \tau}\right]
\qquad ({h} < 0),
\\[\medskipamount]
\Psi_E = (f^{-{1\over 2}}e^{-{1\over 2}i\,\theta}  D^{-1})
&\displaystyle \sum_{k > 0}&
\left[(c_k^+ e^{i k x}\pmatrix{1 \cr -i \cr}
+ c_k^- e^{-i k x}\pmatrix{1 \cr i \cr}) \, e^{ k \tau}\right.
\nonumber\\[\smallskipamount]
&&\left.{}+
(d_k^+ e^{i k x}\pmatrix{1 \cr i \cr}
+ d_k^- e^{-i k x}\pmatrix{1 \cr -i \cr}) \, e^{- k \tau}\right]
\qquad ({h} > 0),
\end{eqnarray}
where $a_k^\pm$, $b_k^\pm$, $c_k^\pm$, and $d_k^\pm$ are arbitrary complex
constants.
(We omit the zero frequency solution.)

\subsection{Asymptotic behavior and continuity of solutions}

Assuming the Fourier sums above are convergent then
the singularity structure of these solutions
in the vicinity of the degeneracy
is determined by
$f^{-{1\over 2}} D^{-1}$:
\begin{equation}
\Psi_{L/E} \simeq \pmatrix{ O(f^{-{1\over 2}} |{h}|^{-{1 \over 4}})\cr
O(f^{-{1\over 2}} |{h}|^{1 \over 4}) \cr}.
\end{equation}
In particular, solutions are bounded if $f \simeq O(|{h}|^{-{1 \over 2}})$.
Thus one cannot have both bounded solutions and a bounded spinor metric at the
degeneracy hypersurface.
One possible choice is $f = |{h}|^{-{1 \over 2}}$, in which case a
continuous
match of a
Lorentzian and Euclidean solution would imply
\begin{equation}
a_k^+ + b_k^+ = c_k^+ + d_k^+, \qquad a_k^- + b_k^- = c_k^- + d_k^-,
\end{equation}
where $\tau(t_0) = 0$ is assumed.  With this choice, requiring
continuity does not induce a bijective map between Lorentzian and Euclidean
solutions.

\section{Currents}\label{sec:currents}

There are two important currents that are locally conserved for solutions
to the massless Dirac equation above.
In regular domains the current
\begin{equation}
j_D^\mu[\Psi,\Xi] = {\mathop{\rm Im\;}\nolimits} (\Psi, \gamma^\mu \Xi)
\label{eq:jD}
\end{equation}
is conserved for solutions $\Psi,\Xi$:
\begin{eqnarray}
\nabla_\mu (\Psi, \gamma^\mu \Xi)
&=& \partial_\mu (\Psi, \gamma^\mu \Xi)
+ {\Gamma^\mu}_{\mu\rho} (\Psi, \gamma^\rho \Xi)
\nonumber\\[\smallskipamount]
&=& (S_\mu \Psi,\gamma^\mu \Xi) + (\Psi,S_\mu (\gamma^\mu \Xi))
+ {\Gamma^\mu}_{\mu\rho} (\Psi, \gamma^\rho \Xi)
\nonumber\\[\smallskipamount]
&=& - (\gamma^\mu S_\mu \Psi,\Xi) + (\Psi,\gamma^\mu S_\mu \Xi),
\end{eqnarray}
using $[S_\mu,\gamma^\nu] = - {\Gamma^\nu}_{\mu\rho} \gamma^\rho$ and
$(\Psi,\gamma^\mu \Xi) = - (\gamma^\mu \Psi,\Xi)$ which follow from the
definitions and properties of the spinor covariant derivative and spinor
metric (see section~\ref{sec:spinor cov}).
For a massless theory the axial vector current is also conserved
\begin{equation}
j_A^\mu[\Psi,\Xi]
= {\mathop{\rm Re\;}\nolimits} (\Psi, {\frak z} \gamma^\mu \Xi),
\end{equation}
where ${\frak z} = \sqrt{|{h}|} \gamma^t\gamma^x$,since
\begin{equation}
\nabla_\mu (\Psi, {\frak z} \gamma^\mu \Xi)
= (\gamma^\mu S_\mu \Psi,{\frak z} \Xi) + (\Psi,{\frak z} \gamma^\mu S_\mu
\Xi).
\end{equation}
Note that $\nabla {\frak z} = 0$, since the connection is metric compatible
and ${\frak z}$ is the metric dual of the metric volume element.

Given
\begin{equation}
\Psi = (f^{-{1\over 2}}e^{-{1\over 2}i\,\theta}  D^{-1}) \, \psi, \qquad
\Xi = (f^{-{1\over 2}}e^{-{1\over 2}i\,\theta}  D^{-1}) \, \xi,
\end{equation}
which are defined piecewise on the non-degenerate parts of $M$, where they
satisfy the massless Dirac equation, we obtain for the components of the Dirac
current
\begin{equation}
j_D^t[\Psi,\Xi] = -|{h}|^{-{1 \over 2}}
{\mathop{\rm Re\;}\nolimits} \psi^\dagger
\pmatrix{1 &  0\cr 0 & -{\mathop{\rm sgn\;}\nolimits} {h}\cr} \xi,
\qquad
j_D^x[\Psi,\Xi]
= {\mathop{\rm Re\;}\nolimits} \psi^\dagger \pmatrix{ 0&  1\cr 1 & 0\cr} \xi,
\end{equation}
and for the components of the axial current
\begin{equation}
j_A^t[\Psi,\Xi] = {h}^{-1}|{h}|^{1 \over 2} {\mathop{\rm Im\;}\nolimits}
\psi^\dagger \pmatrix{ 0&  1\cr 1 & 0\cr} \xi,
\qquad
j_A^x[\Psi,\Xi]
= {\mathop{\rm Im\;}\nolimits} \psi^\dagger
\pmatrix{ 1&  0\cr 0 & -{\mathop{\rm sgn\;}\nolimits} {h}\cr} \xi.
\end{equation}

The continuity of these currents depends on the assumptions made for the
continuity of the spinor components. From our discussion above it is clear
that this requires some assumptions about the behavior of the spinor metric
in the vicinity of the signature change.

However some purely signature dependent effects can be seen by considering
the coordinate independent contractions
\begin{equation}
g_{\mu\nu} j_{D/A}^\mu[\Psi,\Xi] j_{D/A}^\nu[\Psi,\Xi] \simeq O(1),
\end{equation}
which stay bounded near the hypersurface of signature change but contain terms
which depend on ${\mathop{\rm sgn\;}\nolimits} {h}$.  (Note that $\psi \simeq
O(1) \simeq \xi$.)  Thus the currents do not exhibit any divergences which
depend on the choice of spinor metric or on ${h}$, although they can be seen
to be discontinuous in general for any linear prescription relating spinor
data across the hypersurface of signature change.

\section{Conclusion}

We have drawn attention to some of the subtleties involved in discussing
spinor fields in the presence of a smooth metric degeneracy.  By insisting on
interpolating smoothly ($C^1$) between the representations on either side of
the degeneracy, we have been able to derive a number of interesting results.
In particular, we have introduced the notion of a spinor fibration and used
this to give a natural interpolation between the notions of a spinor on the
two
sides of the degeneracy.  This enables one to discuss the concept of
continuity of a spinor field in this context.
Despite the absence of a continuous field of local orthonormal frames
we have shown how a local massless Dirac
equation can be constructed, albeit in terms of a class of spinor metrics
equivalent up to local scalings and a phase freedom associated with charge
conjugation.
We have shown that the singularity structure of the solutions at
metric degeneracies depends on the choice of spinor metric.
An important conclusion of our work is that it is impossible to have both a
continuous spinor metric and continuous solutions to the Dirac equation.
Researchers studying spinor fields on manifolds with smooth degenerate metrics
will be forced to make a choice.
Furthermore, our
formalism allows one to determine explicitly how various assumptions regarding
the continuity of the spinor components affect the continuity of the Dirac
current.

A dynamic theory of spinors on a degenerate background geometry may require a
dynamical prescription to remove the freedom inherent in the construction of
the spinor connection. One way to implement this idea would be to promote the
scaling degree of freedom in the spinor metric to an independent scalar field
and include this in the dynamical theory. A less radical suggestion might be
to relinquish completely the irreducible spinor representations for matter by
embedding a multiplet of spinor fields into a single K\"ahler field.  The
natural dynamics of such a multi-component tensor field depends only on the
metric structure of the manifold which is no longer required to sustain a
spinor structure.

Relinquishing the assumption of a smooth interpolation of representations on
either side of the metric degeneracy may lead to an alternative construction
of a spinor fibration.  However, it is unlikely to circumvent the
discontinuity of the currents which was found to be purely an algebraic effect
of the signature change.

\acknowledgements

TD, CAM, and JS would like to thank the School of Physics \& Chemistry at
Lancaster University for kind hospitality.  Maple was used extensively to
manipulate Clifford algebra expressions.  This work was partially supported by
NSF Grant PHY-9208494 (CAM, TD, \& JS), a Fulbright Grant (CAM), and the Human
Capital and Mobility Programme of the European Union (RWT).  CW is grateful to
the Committee of Vice-Chancellors and Principals for an Overseas Research
Studentship, Lancaster University for a Peel Studentship, and the School of
Physics \& Chemistry at Lancaster University for a School Studentship.

\newcounter{list}
\begin{list}{\hbox{$\fnsymbol{list}$}}
{\labelwidth\WidestRefLabelThusFar  \labelsep4pt %
\leftmargin\labelwidth %
\advance\leftmargin\labelsep
\usecounter{list}}
\item
Permanent address: Dept.\ of Mathematics, Oregon State University, Corvallis,
OR 97331; email: {\tt tevian{\rm @}math.orst.edu}
\item
Permanent address: Dept.\ of Physics, Oregon State University, Corvallis, OR
97331; email: {\tt corinne{\rm @}physics.orst.edu}
\item
email: {\tt J.Schray{\rm @}lancaster.ac.uk}
\item
email: {\tt r.w.tucker{\rm @}lancaster.ac.uk}
\item
email: {\tt cwang{\rm @}lavu.lancs.ac.uk}
\end{list}

\end{document}